\documentclass[aps,twocolumn,prc,superscriptaddress,preprintnumbers,nofootinbib,floatfix,amssymb,amsfonts,amsmath]{revtex4-2}
\usepackage{graphicx}
\usepackage{epsf}
\usepackage{amsmath,amssymb}
\usepackage{bm}
\usepackage{color}
\usepackage{ulem}
\usepackage{dcolumn}

\newcommand{\nuc}[2]{{}^{#2} \mathrm{#1}}
\begin{document}
%
\title{
Cluster-shell competition 
and effect of adding hyperons
} 
\author{Naoyuki Itagaki}
\affiliation{
Department of Physics, Osaka Metropolitan University,
Osaka 558-8585,  Japan
}
\affiliation{
Nambu Institute for Theoretical and Experimental Physics, Osaka Metropolitan University,
Osaka 558-8585,  Japan
}
\author{Emiko Hiyama}
\affiliation{
Department of Physics, Graduate School of Science,
Tohoku University,
Sendai 980-8578, Japan
}
\affiliation{
RIKEN Nishina Center for Accelerator-Based Science,
 Wako 351-0198, Japan
}
%
%
\date{\today}
\begin{abstract}
\begin{description}
\item[Background]
The fundamental question is how the hyperon plays a role in the nuclear structure. It is of particular importance, especially in the light mass region, to verify the structure change when $\Lambda$ particle(s) is added to normal nuclei.
\item[Purpose]
The ground state of $\nuc{Be}{8}$ has been know to have a well-developed $\alpha$--$\alpha$ cluster
structure, whereas
$\nuc{C}{12}$ has a mixed structure of three $\alpha$ clusters and $jj$-coupling shell model,
where $\alpha$ clusters are partially broken.  
Adding $\Lambda$ particle(s) could induce the structure change. We compare the Be and C cases.
\item[Methods]  
 Using  the antisymmetrized quasi-cluster model (AQCM), the $\alpha$-cluster states and $jj$-coupling shell-model states
 of $\nuc{Be}{8}$ and $\nuc{C}{12}$ are prepared on the same footing, and we add $\Lambda$ particles. 
The cluster-shell competition in the ground state can be well described with this model. 
Using AQCM, we calculate $^8$Be,  $^{9}_{\Lambda}$Be, $^{10}_{\Lambda\Lambda}$Be,
$^{12}$C, $^{13}_{\Lambda}$C, and $^{14}_{\Lambda\Lambda}$C.  
\item[Results]
By adding  one or two $\Lambda$ particle(s), the ground state of $\nuc{C}{12}$ approaches the $jj$-coupling shell model side. On the other hand, in the Be case, although the $\Lambda$ particle(s) shrinks the $\alpha$--$\alpha$ distance,
the breaking effect of the cluster structure is rather limited.
\item[Conclusions] 
The spin-orbit interaction is the driving force of breaking the $\alpha$ clusters, and whether the glue-like effect of $\Lambda$ particle(s)
attracts the cluster inside the range of this interaction is crucial.
In $^{14}_{\Lambda\Lambda}$C, the breaking of $\alpha$ clusters in $^{12}$C is much enhanced 
by the addition of the $\Lambda$ particles than the case of free $^{12}$C.
We also found that breaking $\alpha$ clusters in the ground state of $^{14}_{\Lambda\Lambda}$C  affects the excited state with the pure cluster structure. 

\end{description}
\end{abstract}
\maketitle
%
\section{Introduction}
\par
One of the most intriguing phenomena of nuclear structure physics is the competition of 
the shell and cluster structures~\cite{PhysRevC.70.054307}. This is attributed to the effect of the spin-orbit interaction, which strengthens the symmetry of the $jj$-coupling shell model.
It is well known that this  interaction
is vital in explaining
the observed magic numbers of $28$, $50$, $82$, and $126$~\cite{Mayer}.
The spin-orbit interaction also has the effect of breaking clusters~\cite{PhysRevC.70.054307}, where some of 
the strongly correlated nucleons are spatially localized. 
\par
Nevertheless, the $\alpha$ cluster structure is known to be important in the light mass region. 
The Be isotopes are known to have the $\alpha$--$\alpha$ cluster structure;
$\nuc{Be}{8}$ decays into two $\alpha$ clusters, and the molecular-orbital structure of 
valence neutrons appears in the neutron-rich Be isotopes~\cite{PhysRevC.61.044306,
PhysRevC.62.034301,
PhysRevC.65.044302},
which is confirmed by the recent $ab\ initio$ shell-model calculation~\cite{Otsuka-NatureComm}.
This persistence of  the $\alpha$--$\alpha$ cluster structure is owing to the  $\alpha$--$\alpha$ distance, which is about 3--4~fm and large enough compared with the range of the spin-orbit interaction.
\par
In light nuclei, it is considered that these two different pictures (shell and cluster) coexist,
and they compete with each other.
Although the $\alpha$--$\alpha$ cluster structure may persist in $\nuc{Be}{8}$,
when one more $\alpha$ cluster is added, in $\nuc{C}{12}$, 
the interaction among $\alpha$ clusters gets stronger, and the system has a shorter  
$\alpha$--$\alpha$
distance~\cite{KAMIMURA1981456,Uegaki12C}.
In this case, the $\alpha$ clusters are trapped in the interaction range of the spin-orbit interaction.
Although the traditional $\alpha$ cluster model (Brink model)~\cite{Brink} is incapable of treating the spin-orbit interaction, its effect is significant if we allow the breaking of the $\alpha$ clusters.
The ground state of $\nuc{C}{12}$ is found to have a mixed nature of shell and cluster components~\cite{PhysRevC.94.064324,PTP.117.655,PhysRevLett.98.032501}.
On the other hand, the second $0^+$ state of $^{12}$C is well known $\alpha$ clustering state called the Hoyle state.
Since this state is nearby the three-$\alpha$ breakup threshold, the wave function is dilute, and this state has a well-developed $\alpha$ clustering structure.
\par 
It is interesting to investigate how clustering structure is changed when a hyperon such
as a $\Lambda$ particle is injected into $^8$Be and $^{12}$C.
Here it should be noted that there is no Pauli principle between
nucleons and a $\Lambda$, and the $\Lambda N$ interaction is attractive, but
weaker than $NN$ interaction.
Using this property,
some authors studied the structure of $^9_{\Lambda}$Be and $^{13}_{\Lambda}$C from
the viewpoint of dynamical change of the core nuclei, $^8$Be and $^{12}$C, due to
the addition of $\Lambda$ particle.
For instance, Motoba {\it et al.} \cite{Motoba83}, pointed out that
the $\alpha$--$\alpha$ distance in $^9_{\Lambda}$Be was shrunk by about 20~\%
in comparison with that in the $^8$Be core nucleus by $\Lambda$
injection.
In the Carbon isotope,
one of the present authors (E. H.) pointed out that
dynamical change due to the addition of a $\Lambda$ particle is
dependent on the states in the core nucleus of $^{12}$C
within the framework of $3\alpha$ and $3\alpha +\Lambda$ three- and
four-body OCM (orthogonal condition model)~\cite{Hiyama-PTP.97.881}.
The ground state of $^{12}$C, $0^+_1$, is a mixture of shell and cluster structure;
the $\alpha$--$\alpha$ distance does not change due to the addition of a $\Lambda$ particle.
On the other hand, the $\alpha$--$\alpha$ distance is dramatically contracted
in the Hoyle state of
$^{13}_{\Lambda}$C, which is well-developed clustering state \cite{Hiyama-PTP.97.881}.
However, it should be noted that
this calculation was done without taking into account the breaking effect of $\alpha$ clusters in
$^{12}$C.
In addition, in Ref.~\cite{FUNAKI2017336}, they discussed 
the similarity and difference in several states of $^{12}$C and $^{13}_{\Lambda}$C.
In this way, there are some discussions on the change of the $\alpha$--$\alpha$
distance w/o the $\Lambda$ particle and the change of the structure.
However, there remain never discussed effects of the clustering in such Be and C isotopes
due to addition of $\Lambda$ particles.
The question is how the clustering is broken when $\Lambda$ particles 
shrinks the $\alpha$--$\alpha$ distance. The traditional cluster model is incapable of describing
such breaking situation and we must extend the model space to incorporate the spin-orbit contribution,
which is the driving force of breaking clusters.
\par
Thus, in this work,  
we focus on how 
the clustering is changed and broken
due to the addition of a $\Lambda$ particle(s)
in 
$^{8}$Be,
$^{9}_{\Lambda}$Be, $^{10}_{{\Lambda \Lambda}}$Be,
$^{12}$C,
$^{13}_{\Lambda}$C, and $^{14}_{{\Lambda \Lambda}}$C. 
In the case of Be isotopes, as mentioned, the $\Lambda$ particle(s) shrinks the $\alpha$--$\alpha$
relative distance~\cite{Motoba-PTP.70.189,Hiyama-PTP.97.881}, 
but the resultant distance might still be outside the range
of the spin-orbit interaction, and the $\alpha$ cluster structure could persist. On the contrary,
when $\Lambda$ particle(s) is added to $\nuc{C}{12}$, the  distances between clusters get
even shorter. Since the spin-orbit interaction works in the inner regions of the nuclear systems,
the breaking of $\alpha$ clusters is expected to be enhanced. Therefore, the ground state would
approach more $jj$-coupling shell-model side. 
Indeed, as shown in the study of antisymmetrized molecular dynamics~\cite{PhysRevC.83.044323},
the slightly deformed ground state of $\nuc{C}{12}$ is changed into a spherical shape in $^{13}_\Lambda$C.
It is worthwhile to check this point in terms of the cluster-shell competition.
\par
In most of the conventional $\alpha$ cluster models,  
the contribution of the non-central interactions (spin-orbit and tensor interactions) vanishes.
To include the spin-orbit effect,
we have developed the antisymmetrized quasi-cluster model
(AQCM)~\cite{PhysRevC.71.064307,PhysRevC.75.054309,PhysRevC.79.034308,PhysRevC.83.014302,PhysRevC.87.054334,ptep093D01,ptep063D01,ptepptx161,PhysRevC.94.064324,PhysRevC.97.014307,PhysRevC.98.044306,PhysRevC.101.034304,PhysRevC.102.024332,PhysRevC.103.044303,PhysRevC.105.024304}.
This method allows us to smoothly transform $\alpha$-cluster model wave functions to
$jj$-coupling shell model ones, and
we call the clusters that feel the effect of the spin-orbit interaction quasi-clusters.
We have previously introduced AQCM to $\nuc{C}{12}$ and discussed the
competition between the cluster states and $jj$-coupling shell model state~\cite{PhysRevC.94.064324}.
The consistent description of $\nuc{C}{12}$ and $\nuc{O}{16}$, which has been a long-standing problem
of microscopic cluster models, has been achieved.
Also, not only the competition between the
cluster states and the lowest shell-model configuration,
the effect of single-particle excitation was further included 
in the description of the ground state~\cite{PhysRevC.103.044303}.
\par
This paper is organized as follows. 
The framework is described in  Sec.~\ref{Frame}.
The results are shown in Sec.~\ref{Results}.
The conclusions are presented in Sec.~\ref{Concl}.
%

\section{framework}
\label{Frame}
%
\par
The wave function is fully antisymmetrized, and
different basis states are superposed based on
the generator coordinate method  (GCM)
after the
angular momentum projection, and 
the amplitude for each basis state is determined by diagonalizing the norm and Hamiltonian matrices.  
\subsection{Single-particle wave function}
\par
In our framework, every single particle is described in a Gaussian form
as in many traditional cluster models, including the Brink model~\cite{Brink},
\begin{equation}	
  \phi^{\tau, \sigma} \left( \bm{r} \right)
  =
  \left(  \frac{2\nu}{\pi} \right)^{\frac{3}{4}} 
  \exp \left[- \nu \left(\bm{r} - \bm{\zeta} \right)^{2} \right] \chi^{\tau,\sigma}, 
  \label{spwf} 
\end{equation}
where the Gaussian center parameter $\bm{\zeta}$
is related to the expectation 
value of the position of the nucleon,
and $\chi^{\tau,\sigma}$ is the spin-isospin part of the wave function.
The $\alpha$ cluster is expressed by four nucleons with different spin and isospin sharing the same
$\bm{\zeta}$ value.
For the size parameter $\nu$, 
here we use $\nu = 1/2b^2$ and $b=1.46$~fm.
The Slater determinant is constructed from 
these single-particle wave functions by antisymmetrizing them.
The $\Lambda$ particle is represented by the same local Gaussian-type wave function.

\par
This traditional $\alpha$ cluster wave function cannot take into account
the effect of non-central interactions including the spin-orbit interaction. 
We can extend the model based on the AQCM,
by which the contribution of the spin-orbit interaction
due to the breaking of $\alpha$ clusters is included.
Here
the $\bm{\zeta}$ values in Eq.~\eqref{spwf} are changed to complex numbers.
When the original value of the Gaussian center parameter $\bm{\zeta}$
is $\bm{R}$,
which is 
real and
related to the spatial position of this nucleon, 
it is transformed 
by adding the imaginary part as
\begin{equation}
  \bm{\zeta} = \bm{R} + i \lambda \bm{e}^{\text{spin}} \times \bm{R}, 
  \label{AQCM}
\end{equation}
where $\bm{e}^{\text{spin}}$ is a unit vector for the intrinsic-spin orientation of this
nucleon. 
The control parameter $\lambda$ is associated with the breaking of the cluster.
After this transformation, the $\alpha$ clusters are called quasi-clusters.
The two nucleons in the same quasi-cluster with opposite spin orientation have $\bm{\zeta}$ values
that are complex conjugate to each other.
This situation corresponds to the time-reversal motion of two nucleons.
\par
In our previous analysis on $\nuc{C}{12}$~\cite{PhysRevC.94.064324}, 
we have 
introduced two parameters representing the distances between quasi-clusters 
and their breaking ($\lambda$).
The
subclosure configuration of 
$ \left( s_{1/2} \right)^4 \, \left( p_{3/2} \right)^8$
of the $jj$-coupling shell model can be obtained at the limit of small relative distances and $\lambda = 1$.

\subsection{Angular momentum projection and GCM}
\par
Each AQCM Slater determinant is projected to the eigenstates of parity and angular momentum by
using the projection operator $P_{J^\pi}^K$,
\begin{equation}
  P_{J^\pi}^K
  =
  P^\pi \frac{2J+1}{8\pi^2}
  \int d\Omega \, {D_{MK}^J}^* R \left(\Omega \right).
\end{equation}
Here ${D_{MK}^J}$ is the Wigner $D$-function 
and $R\left(\Omega\right)$ is the rotation operator
for the spatial and spin parts of the wave function.
This integration over the Euler angle $\Omega$ is numerically performed.
The operator $P^\pi$ is for the parity projection ($P^\pi = \left(1+P^r\right) / \sqrt{2}$ for
the positive-parity states, where $P^r$ is the parity-inversion operator), 
which is also performed numerically.
\par
The AQCM basis states with different distances between quasi-clusters and $\lambda$ values
are superposed based on GCM.
We also generate Gaussian centers for the $\Lambda$ particles using random numbers, 
and the basis states with different positions are superposed.
The coefficients $\left\{ c^K_i \right\}$ for the linear combination of the Slater determinants
are obtained together with the energy eigenvalue $E$
when we diagonalize the norm and Hamiltonian matrices, namely  
by solving the Hill-Wheeler equation.
 \begin{equation}
  \sum_{j} (<\Phi_i | (P_{J^\pi}^{K})^\dagger H P_{J^\pi}^K | \Phi_j>  - E <\Phi_i | (P_{J^\pi}^{K})^\dagger P_{J^\pi}^K | \Phi_j>)c^K_j = 0.
 \end{equation}
\subsection{Hamiltonian}
\par
The Hamiltonian consists of kinetic energy and 
potential energy terms.
For the potential part, the interaction consists of the central, spin-orbit, and Coulomb terms. 
The nucleon-nucleon interaction is Volkov No.2~\cite{VOLKOV196533} with the Majorana exchange parameter
of $M = 0.6$, which has been known to reproduce the scattering phase shift of $^4$He--$^4$He~\cite{PTP.61.1049}.
For the spin-orbit part, we use the spin-orbit term of the G3RS interaction~\cite{PTP.39.91},
which is a realistic interaction originally developed to reproduce the nucleon-nucleon scattering phase shifts.
The strength of the spin-orbit interactions~\cite{PhysRevC.94.064324}
 is set to $V_{ls}^1=V_{ls}^2=1450 \, \mathrm{MeV}$,
which reproduces the binding energy of $^{12}$C from the three-$\alpha$ threshold.
For the nucleon-$\Lambda$ interaction, we employ only the central part;
YNG-ND interaction~\cite{Yamamoto-ptp.117.361}. The $k_F$ value for $^9_\Lambda$Be and $^{10}_{\Lambda \Lambda}$Be
is 0.962~fm$^{-1}$ as in Ref.~\cite{Hiyama-PTP.97.881} and 1.17~fm$^{-1}$ for $^{13}_\Lambda$C and $^{14}_{\Lambda \Lambda}$C
as in Ref.~\cite{PhysRevC.83.044323}.
For the $\Lambda$-$\Lambda$ interaction, we adopt the 
one called ``NS'' in Ref.~\cite{Yamamoto-ptp.117.361}, which allows the reproduction of the
binding energy of $^{6}_{\Lambda \Lambda}$He.

\section{Results}
\label{Results}

\begin{figure}[tb]
  \centering
  \includegraphics[width=5.5cm]{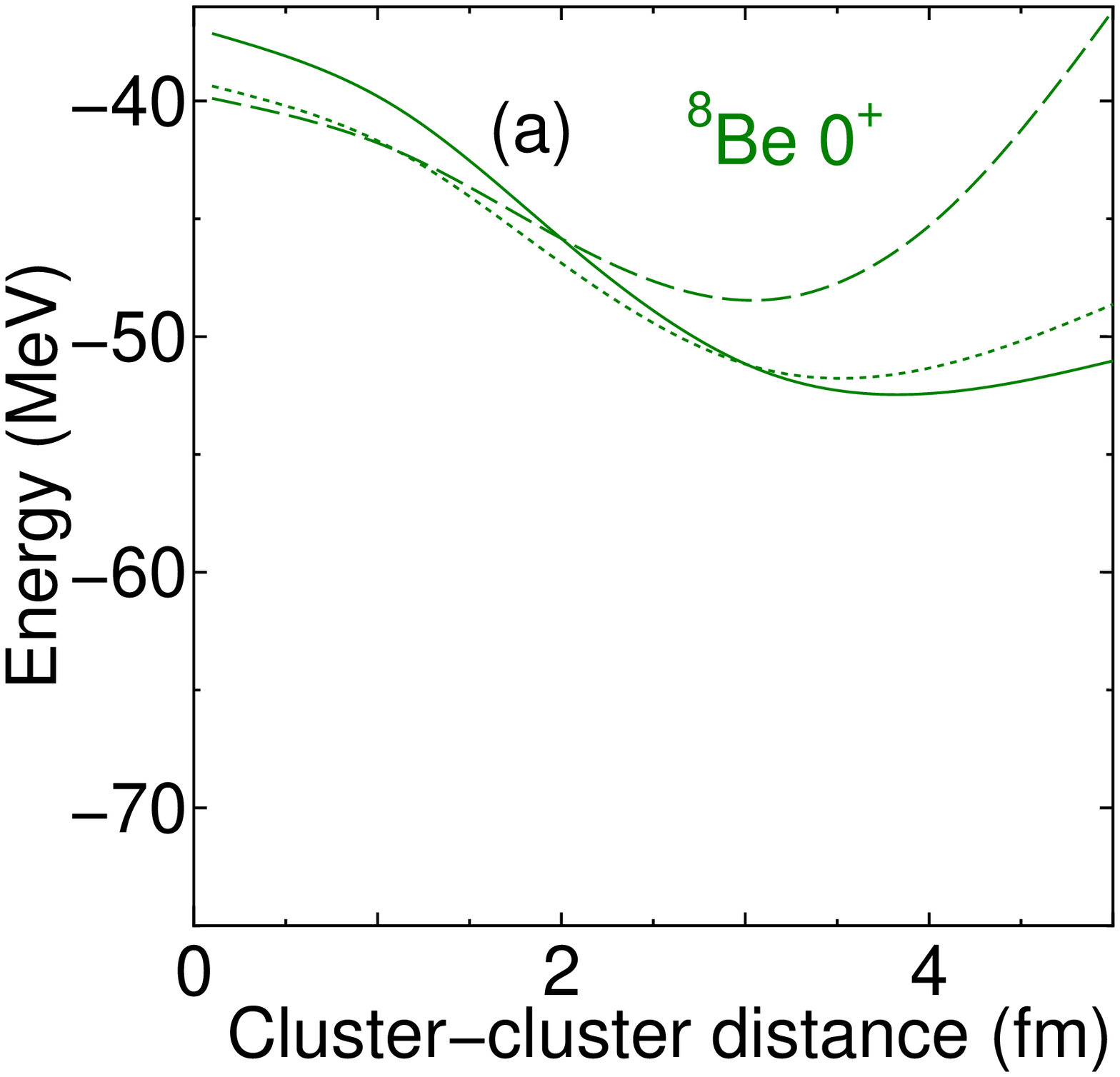} 
    \includegraphics[width=5.5cm]{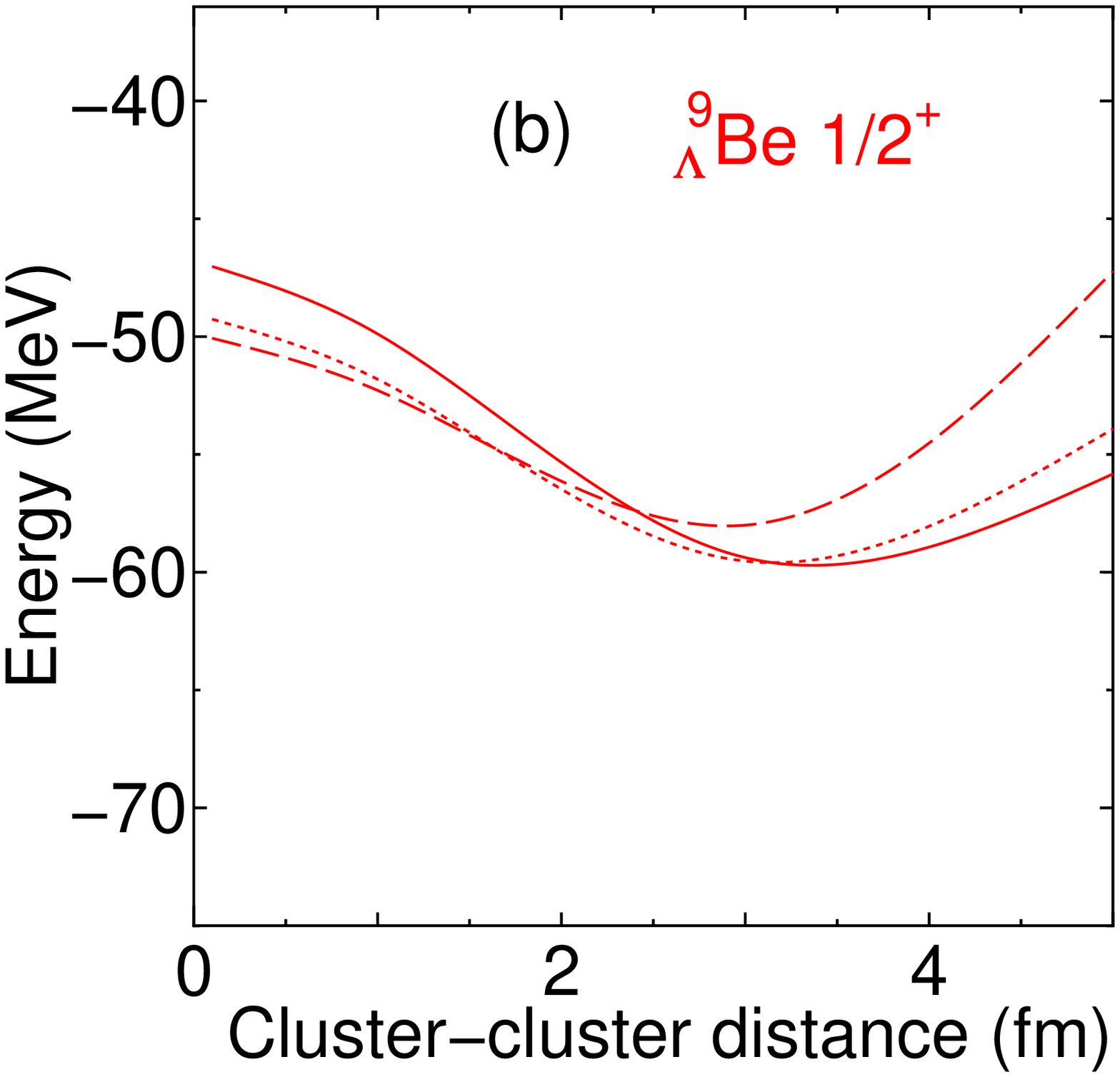} 
      \includegraphics[width=5.5cm]{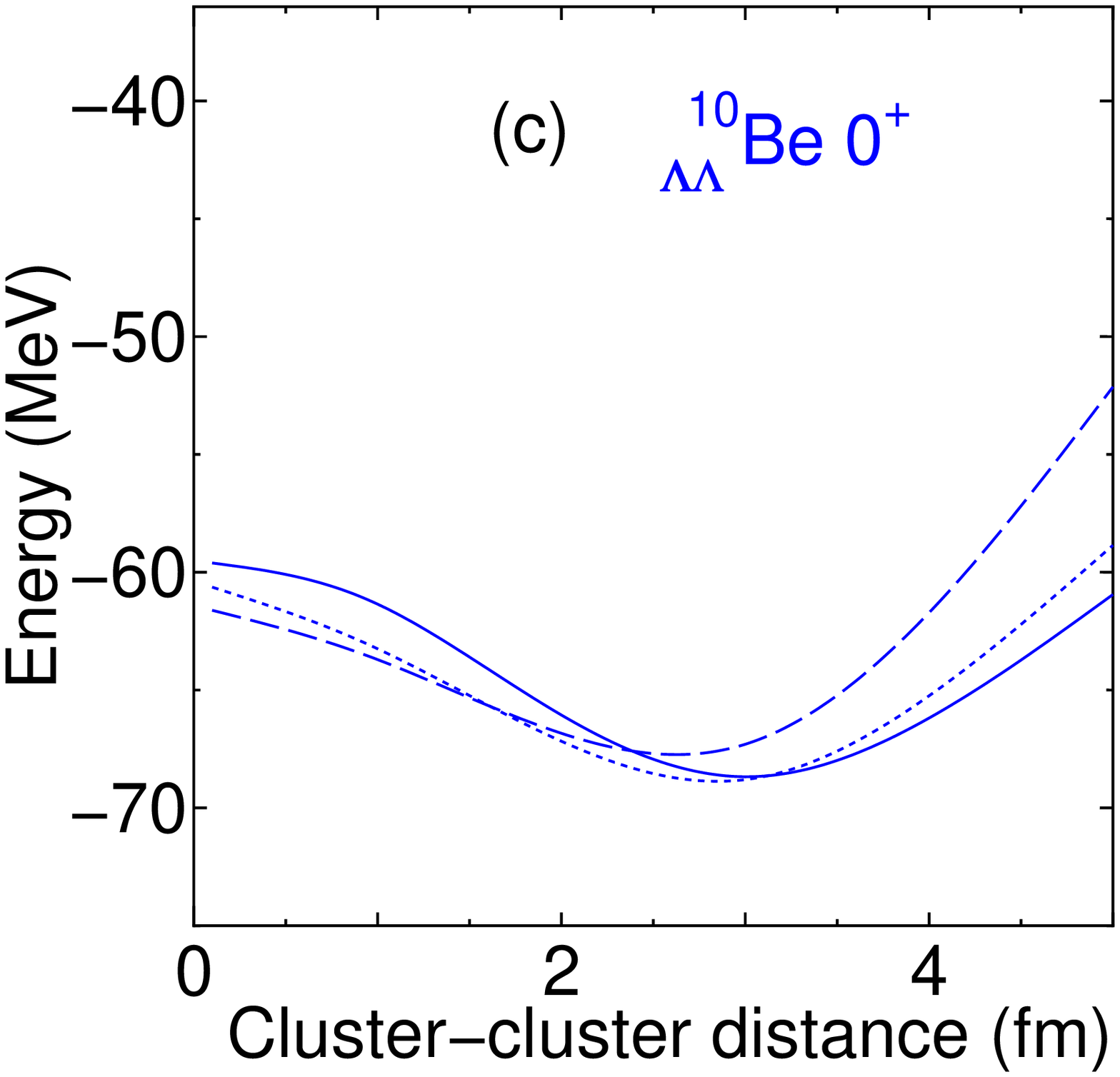} 
  \caption{
(a): Energy curves of $0^+$ state of 
$\nuc{Be}{8}$ as a function of the distance between two $\nuc{He}{4}$ clusters. Solid line is for $\lambda = 0$ (pure two $\alpha$'s)
and dotted and dashed lines are for two quasi-clusters with  $\lambda=0.1$ and 0.2, respectively. (b): Same as (a) but for 
the $1/2^+$ state of $^9_\Lambda$Be. (c): Same as (a) but for the $0^+$ state of $^{10}_{\Lambda \Lambda}$Be. 
    }
  \label{be8-be9L-be10LL}
\end{figure}

\subsection{Ground states of $\nuc{Be}{8}$, $^9_\Lambda$Be, and $^{10}_{\Lambda \Lambda}$Be}
\par
We start the discussion with $\nuc{Be}{8}$.
Our Hamiltonian gives the energy of $-27.57$~MeV for the $\alpha$ cluster, and thus,
$-55.1$~MeV is the two-$\alpha$ threshold energy (experimentally $-56.6$~MeV, to which our theoretical value does not contradict).
Figure~\ref{be8-be9L-be10LL}~(a) shows the
energy curves of the $0^+$ state of 
$\nuc{Be}{8}$ as a function of the distance between two $\nuc{He}{4}$ clusters. The solid line is for $\lambda = 0$ (pure two $\alpha$'s),
and the dotted and dashed line are for two quasi-clusters with $\lambda=0.1$ and 0.2, respectively. 
The energy minimum point appears around the relative distance of $\sim$3.5~fm. 
This distance is quite large, and this is outside of the interaction range of the spin-orbit interaction.
Therefore, the $\lambda$ value that gives the minimum energy is zero (solid line),
which means that the  $\alpha$ clusters are not broken.
The $\alpha$ breaking effect can be seen in more inner regions, where the energies of dotted and dashed lines
are lower than the solid line.
The $\alpha$ clusters are surely broken there. 
However, at short relative distances,
the energy  itself is high enough, and the spin-orbit interaction only plays
a role in reducing the increase of the excitation energy to some extent
when two clusters get closer.
\par
The situation is slightly different in Figure~\ref{be8-be9L-be10LL}~(b), which is  for the $1/2^+$ of $^9_\Lambda$Be,
where one $\Lambda$ particle is added.
We superpose 50 Slater determinants with different positions for the $\Lambda$ particle 
and diagonalize the Hamiltonian based on the GCM
for each cluster-cluster distance and $\lambda$.
Owing to the $\Lambda$ particle added, the attractive effect is increased, and the optimal distance
between the two $\nuc{He}{4}$ nuclei (lowest energy point) is around 3~fm,
slightly shorter than the $\nuc{Be}{8}$ case.
Here, the solid line ($\lambda = 0$) and the dotted line ($\lambda = 0.1$) almost degenerate, 
and thus, the $\alpha$ clusters are slightly broken due to the spin-orbit effect.
The tendency is a bit enhanced in $^{10}_{\Lambda \Lambda}$Be shown in Fig~\ref{be8-be9L-be10LL}~(c).
The optimal cluster-cluster distance is less than 3~fm, where the dotted line ($\lambda = 0.1$) is slightly lower than
the solid line ($\lambda = 0$). 
The number of Slater determinants with different positions for the $\Lambda$ particles 
is increased to 100 for each $^4$He--$^4$He distance and $\lambda$.
In this way, since the $^4$He--$^4$He distances are large in $^9_{\Lambda}$Be
and $^{10}_{\Lambda \Lambda}$Be, we find that
the $\alpha$-cluster braking effect is rather small.

\subsection{Ground states of $\nuc{C}{12}$, $^{13}_\Lambda$C, and $^{14}_{\Lambda \Lambda}$C}
\par
\begin{figure}[tb]
  \centering
  \includegraphics[width=5.5cm]{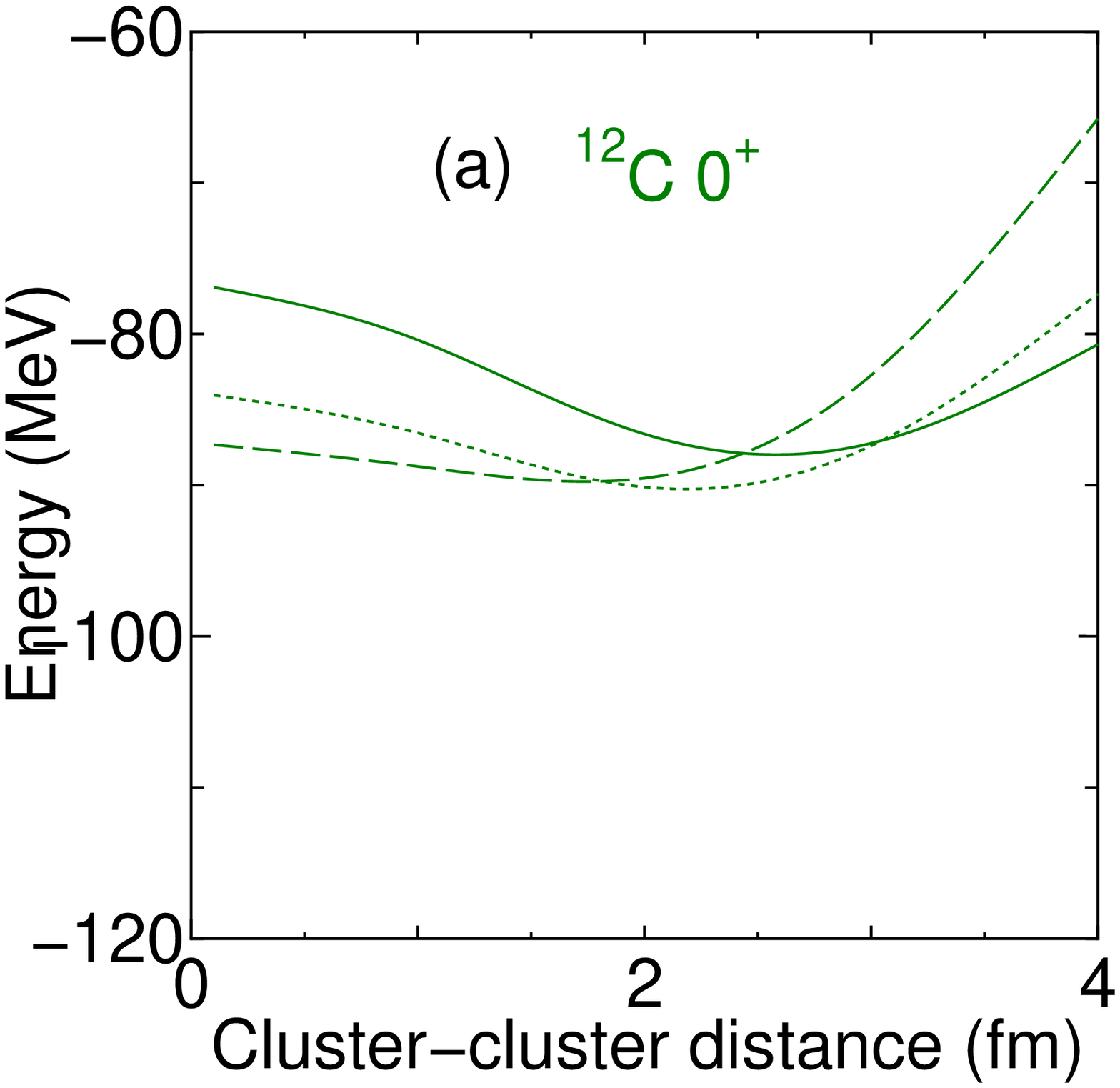} 
   \includegraphics[width=5.5cm]{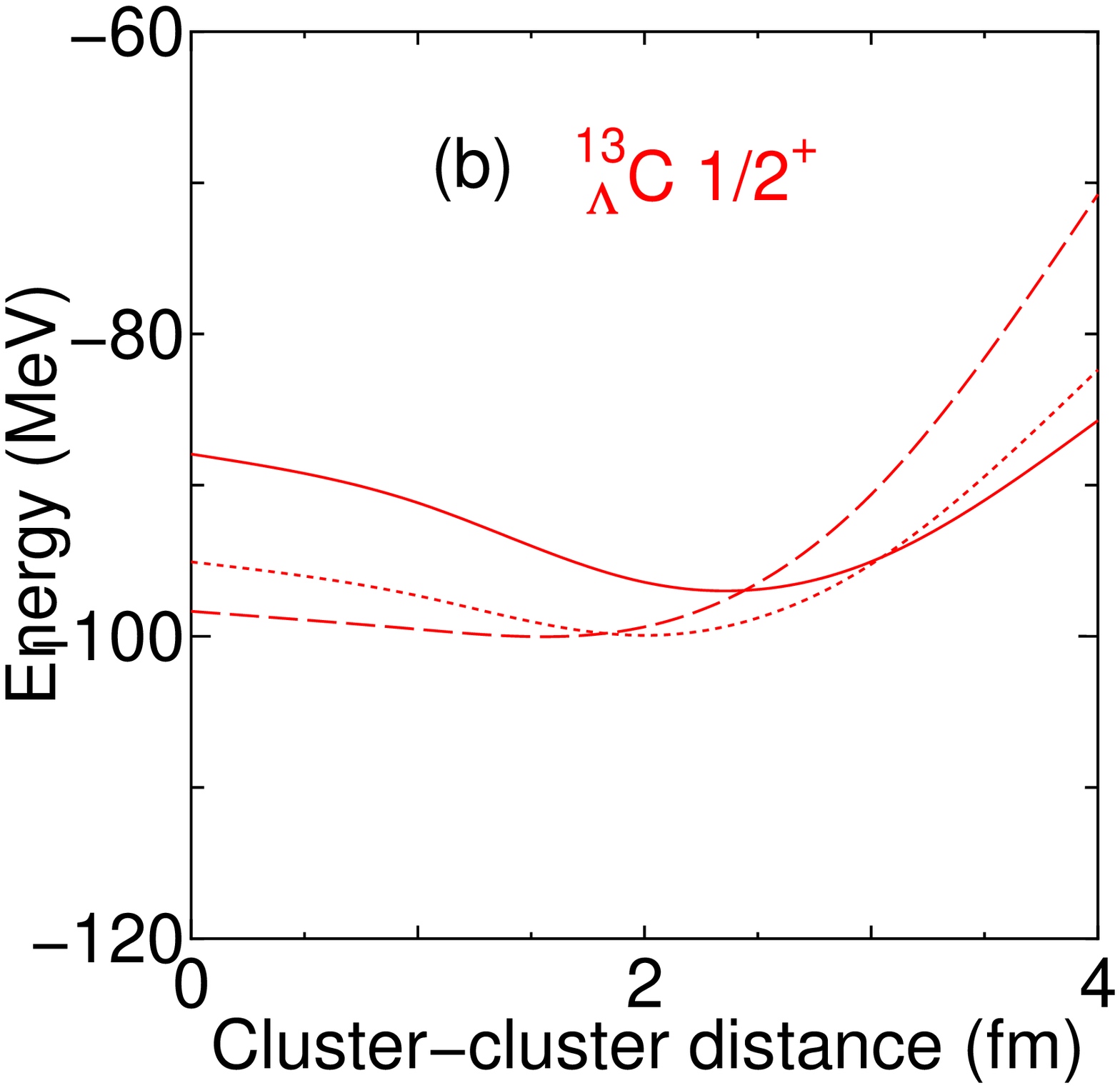} 
    \includegraphics[width=5.5cm]{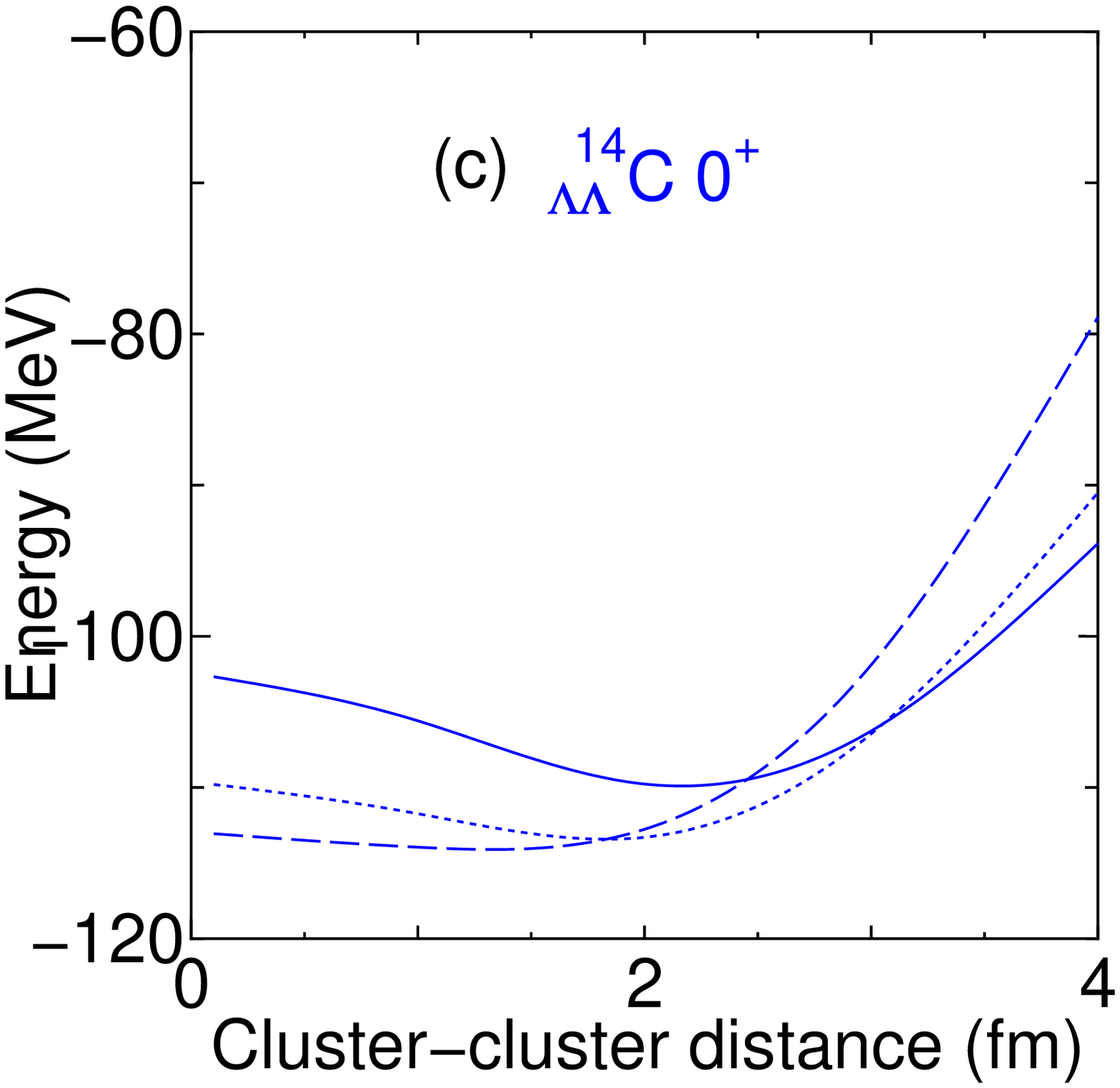} 
\caption{
(a): Energy curves of $0^+$ state of 
$\nuc{C}{12}$ as a function of the distance between three $\nuc{He}{4}$ clusters
with equilateral triangular configuration.
Solid line is for $\lambda = 0$ (pure three $\alpha$'s)
and dotted and dashed lines are for two quasi-clusters with  $\lambda=0.1$ and 0.2, respectively.
(b): Same as (a) but for 
the $1/2^+$ state of $^{13}_\Lambda$C. (c) Same as (a) but for 
the $0^+$ state of $^{14}_{\Lambda \Lambda}$C
    }
  \label{c12-c13L}
\end{figure}

Next we discuss $\nuc{C}{12}$ and $^{13}_\Lambda$C, and $^{14}_{\Lambda \Lambda}$C.
The three-$\alpha$ threshold energy is $-82.7$~MeV in our calculation compared with the 
experimental value of $-84.9$~MeV.
Figure~\ref{c12-c13L}~(a) shows the
energy curves of $0^+$ state of 
$\nuc{C}{12}$ with an equilateral triangular configuration
as a function of the distance between two $\nuc{He}{4}$ clusters. The solid line is for $\lambda = 0$ (pure three $\alpha$'s).
Since one $\nuc{He}{4}$ is added to $\nuc{Be}{8}$,
the energy minimum point appears around the relative distance of 2.5--3.0~fm, shorter by
1~fm than the previous $\nuc{Be}{8}$ case before allowing the breaking of $\alpha$ clusters.
Therefore, it is considered that the three $\alpha$ clusters step in the interaction range of the spin-orbit interaction.
The dotted line ($\lambda=0.1$) and dashed line ($\lambda =0.2$) almost degenerate at the
region of the lowest energy (the relative cluster-cluster distance shrinks to 2~fm there).
\par
This tendency is enhanced in Fig.~\ref{c12-c13L}~(b), which is  for the $1/2^+$ of $^{13}_\Lambda$C,
where one $\Lambda$ particle is added.
Owing to the $\Lambda$ particle added, the attractive effect is increased, and the optimal distance
between the $\nuc{He}{4}$ nuclei is around 2.5~fm (solid line) before breaking the $\alpha$ clusters.
When we allow the breaking, the energy curves become almost flat inside the
relative $^4$He--$^4$He distance of 2~fm. The energy minimum points of the dotted ($\lambda = 0.1$) and dashed ($\lambda = 0.2$) lines are
lower than that of the solid line ($\lambda = 0$).
\par
The attractive effect of the $\Lambda$ particles is much more enhanced in Fig.~\ref{c12-c13L}~(c), which is  for the $0^+$ state of $^{14}_{\Lambda \Lambda}$C.
The optimal distance
between the $\nuc{He}{4}$ nuclei (energy minimum point) is around 2.2~fm before breaking the $\alpha$ clusters (solid line).
When we allow the breaking, the energy minimum point appears at the
relative cluster--cluster distance of $\sim$1.4~fm, where the dashed line ($\lambda$=0.2) gives the lowest energy,
and $\alpha$ clusters are significantly broken.
We can confirm that the optimal cluster distance gets shorter, and the breaking of $\alpha$ clusters becomes larger
with the increasing number of $\Lambda$ particles added to the system.

%
%

\subsection{Superposition of states with different $^4$He$-^4$He distance and breaking parameter $\lambda$}
\par
To demonstrate the relation between the effect of $\alpha$ breaking and spin-orbit interaction, we calculate the ground state energies of $^8$Be, $^9_{\Lambda}$Be,
$^{10}_{\Lambda \Lambda}$Be (Table~\ref{obls-Be}) and those of $^{12}$C, $^{13}_{\Lambda}$C, 
$^{14}_{\Lambda \Lambda}$C  (Table~\ref{obls-C}) with two models: ``AQCM''' which explicitly takes account of the breaking effect of 
$\alpha$, and ``Brink model'' which does not involve the $\alpha$ breaking effect ($\lambda = 0$).
We superpose Slater determinants with different positions of the $\Lambda$ particle(s), 
$^4$He--$^4$He cluster distances, 
and $\alpha$-breaking parameter $\lambda$
and diagonalize the Hamiltonian based on the GCM. 
\par
For the Be case (Table~\ref{obls-Be}), 
the energy difference between Brink and AQCM is less than 0.2~MeV in $\nuc{Be}{8}$,
which means that the spin-orbit interaction does not break the $\alpha$ clusters
since they are separated by a certain distance.
The situation is basically the same when $\Lambda$ particle(s) is added.
The difference is about 0.5-0.6 MeV in  $^{9}_{\Lambda}$Be and  $^{10}_{\Lambda\Lambda}$Be.
Concerning the ground state energy of $^{10}_{\Lambda \Lambda}$Be,
the binding energy ($B_{\Lambda\Lambda}$) of $17.5\pm0.4$~MeV from $^8$Be
 has been reported in Ref.~\cite{PhysRevLett.11.29},
which has been revised to $14.7\pm0.4$~MeV in Ref.~\cite{Dalitz-1989} (see the discussions 
in Refs.~\cite{PhysRevC.66.024007,annurev-nucl-101917-021108}),
and the present result (15.23~MeV) is almost consistent with the latter case.

\par
For the C case (Table~\ref{obls-C}),
the energy difference between Brink and AQCM is about 3.3~MeV in $\nuc{C}{12}$, 
and this is much enhanced with the increasing number of the $\Lambda$ particles added. 
The difference increases to 5.2 MeV in $^{14}_{\Lambda \Lambda}$C.
This is because the spin-orbit interaction works in the inner region of the nuclear systems; the glue-like effect of $\Lambda$ particles shrinks the system and induces more contribution of the spin-orbit interaction.
\par
To clarify the mixing of the $jj$-coupling shell model components in each state, 
we utilize the expectation value of the
one-body spin-orbit operator,
\begin{equation}
  \hat{O}^{LS} = \sum_i \bm{l}_i \cdot \bm{s}_i /\hbar^2, 
\end{equation}
where $\bm{l}_i$ and $\bm{s}_i$ are the orbital angular momentum and the spin operators 
for the $i$th nucleon. The sum runs over the nucleons.
The expectation value is zero for the pure $\alpha$ cluster state 
owing to the antisymmetrization effect.
Also,
the $\bm{l}_i \cdot \bm{s}_i /\hbar^2$ value  
is $0.5$ for one nucleon in the $p_{3/2}$ orbit, and the eigen value
is $4$ for the subclosure configuration
of the $jj$-coupling shell model
$(\left(s_{1/2}\right)^4\,\left(p_{3/2}\right)^8)$ in $^{12}$C.
\par
The expectation values of the one-body spin-orbit operator for the ground states of
$\nuc{Be}{8}$, $^{9}_\Lambda$Be, 
and $^{10}_{\Lambda \Lambda}$Be are listed in the column
``one-body LS'' in Table~\ref{obls-Be}. Although the value increases with the number of
$\Lambda$ particles added, it is rather small and cluster structure is considered to be not broken.
However,  this is completely different in the C case. 
The expectation values of the one-body spin-orbit operator for the ground states of
$\nuc{C}{12}$, $^{13}_\Lambda$C, 
and $^{14}_{\Lambda \Lambda}$C are listed in the column
``one-body LS'' in Table~\ref{obls-C}.
The value is $1.55$ for $\nuc{C}{12}$,
and we can reconfirm that the ground state has mixed configurations of shell and cluster aspects.
As the number of the $\Lambda$ particles added increases, we can see that the ground states
approach the $jj$-coupling shell model side.
The values for 
$^{13}_\Lambda$C
and $^{14}_{\Lambda \Lambda}$C
are 1.86 and 2.05, respectively.

\begin{table}
\caption{
Ground state energies of $\nuc{Be}{8}$, $^{9}_\Lambda$Be, 
and $^{10}_{\Lambda \Lambda}$Be (``energy ($J^\pi$)'')
after performing the GCM calculations.
``Brink'' is for the Brink model  ($\lambda =0$); 
two-$\alpha$ clusters without the breaking,
and ``AQCM'' is for the AQCM calculation, where different $\lambda$ states are mixed.
``one-body LS'' is for the expectation values of the one-body spin-orbit operator. 
The values in the parenthesis show the experimental values.
$B_\Lambda$, $B_{\Lambda\Lambda}$ are also shown.
All energies are in MeV.
}
  \begin{tabular}{cccc} 
 \hline
   $^{8}$Be  &  energy ($0^+$) &  &one-body LS \\
  \hline
   Brink        &       $-54.75$     &      &0.00  \\
   AQCM      &       $-54.94$  & ($-56.50 $)     &  0.12        \\
\hline
 $^{9}_\Lambda$Be & energy ($1/2^+$) & $B_\Lambda$ & one-body LS \\
 \hline
   Brink         &       $-60.97$                    &  &  0.00 \\
   AQCM       &       $-61.53$     &6.59 (6.71~\cite{PhysRevC.83.044323})  &  0.29  \\
\hline
   $^{10}_{\Lambda\Lambda}$Be & energy ($0^+$) & $B_{\Lambda\Lambda}$ & one-body LS \\
 \hline
   Brink          &      $-69.60$   &     &   0.00 \\
   AQCM        &      $-70.17$   & 15.23 ($14.7\pm0.4$~\cite{Dalitz-1989}) &    0.44 \\
\hline
\end{tabular}
\label{obls-Be}
\end{table}

\begin{table}
\caption{
Ground state energies of $\nuc{C}{12}$, $^{13}_\Lambda$C, 
and $^{14}_{\Lambda \Lambda}$C  (``energy ($J^\pi$)'')
after performing the GCM calculations.
``Brink'' is for the Brink model  ($\lambda =0$); 
three-$\alpha$ clusters with equilateral triangular shapes without the breaking,
and ``AQCM'' is for the AQCM calculation, where different $\lambda$ states are mixed.
``one-body LS'' is for the expectation values of the one-body spin-orbit operator. 
The values in the parenthesis show the experimental values.
All energies are in MeV.
}
  \begin{tabular}{cccc} 
 \hline
   $^{12}$C  &  energy ($0^+$) &  & one-body LS \\
  \hline
   Brink        &       $-86.84$                  &   & 0.00       \\
   AQCM      &       $-90.12$   & ($-92.16 $)   & 1.55        \\
\hline
 $^{13}_\Lambda$C & energy ($1/2^+$)& $B_\Lambda$  & one-body LS \\
 \hline
   Brink         &       $-97.77$                      &                     &  0.00 \\
   AQCM       &       $-102.00$    & 11.88 (11.69~\cite{PhysRevC.83.044323}) &  1.86  \\
\hline
   $^{14}_{\Lambda\Lambda}$C & energy ($0^+$) & $B_{\Lambda\Lambda}$ & one-body LS \\
 \hline
   Brink          &      $-110.58$  &   &  0.00 \\
   AQCM        &      $-115.74$  & 25.62  &  2.05 \\
\hline
\end{tabular}
\label{obls-C}
\end{table}

\begin{figure}[tb]
  \centering
  \includegraphics[width=5.5cm]{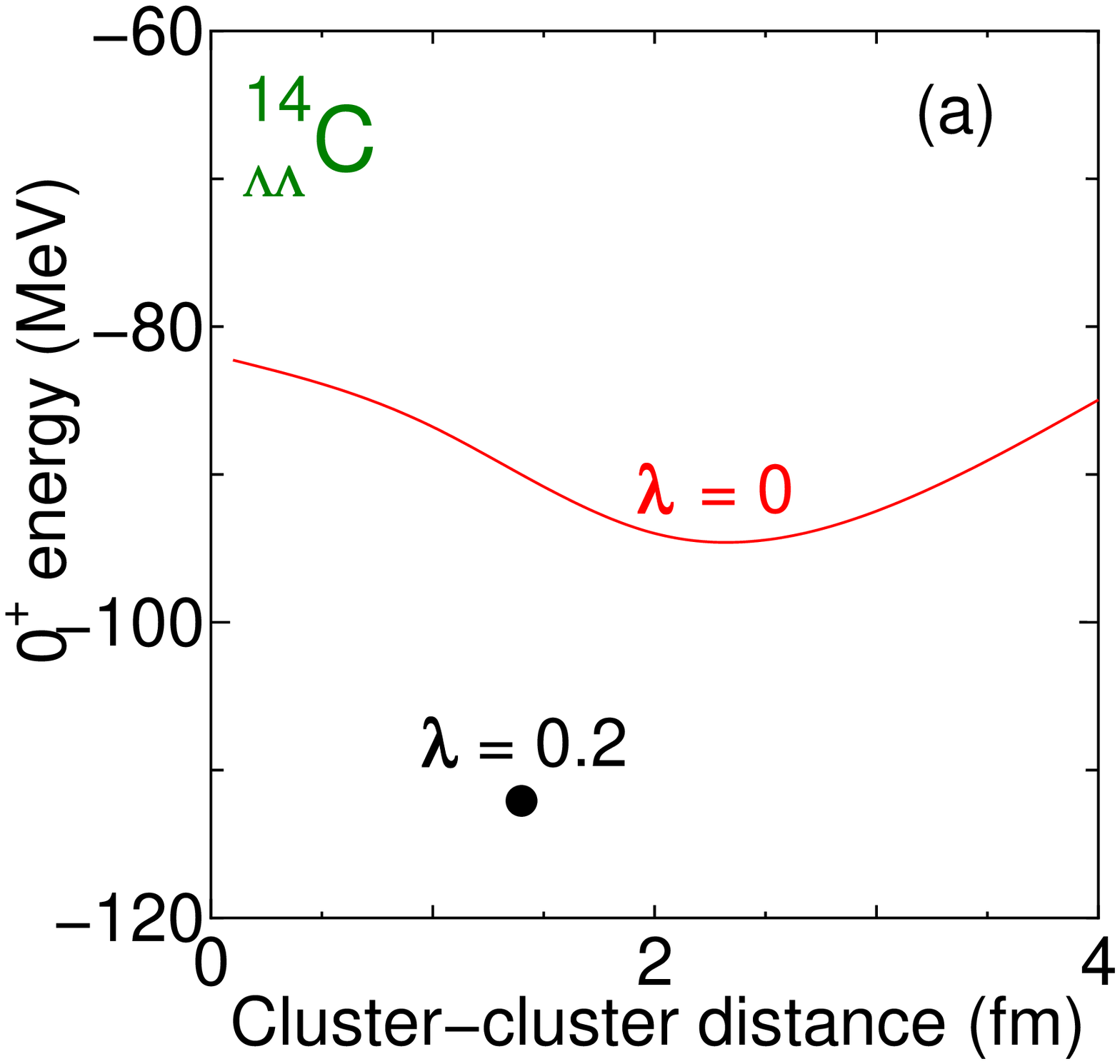} 
  \includegraphics[width=5.5cm]{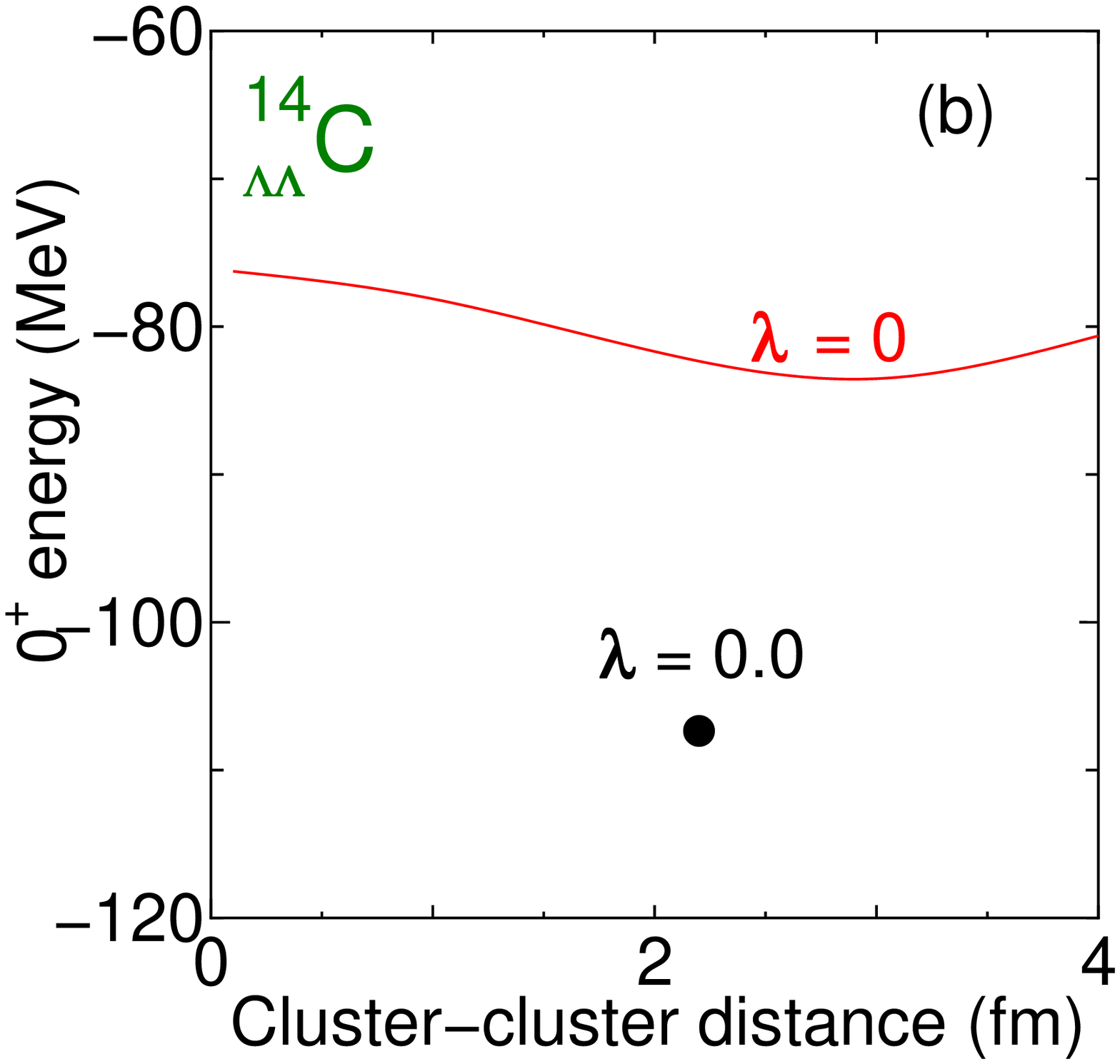} 
  \caption{
Excited $0^+$ state comprised of pure three $\alpha$ clusters in $^{14}_{\Lambda \Lambda}$C
as a function of distances between $\alpha$--$\alpha$ (solid lines).  Ground state is represented by the AQCM
basis state with  the $^4$He--$^4$He distance of $1.4$~fm and $\lambda = 0.2$ (a) and $^4$He--$^4$He distance of $2.2$~fm and $\lambda = 0.0$ (b),
which are shown by the solid circles.
}
  \label{pure-cluster}
\end{figure}
%

\subsection{pure $\alpha$ cluster state orthogonal to the ground state}
\par
We have discussed that the ground states shift to the $jj$-coupling shell model side
by adding $\Lambda$ particles, and 
the final question is where the ``pure'' three-$\alpha$ cluster state appears in $^{14}_{\Lambda \Lambda}$C.
We can discuss it by preparing the pure three-$\alpha$ cluster states and orthogonalizing them to the ground state.
The shift of the ground state to the $jj$-coupling shell-model-side after 
allowing the breaking of $\alpha$ clusters  is found to play a crucial role.
\par 
The solid line in Fig.~\ref{pure-cluster}~(a) shows the excited $0^+$ state
with equilateral triangular configurations of pure three-$\alpha$ clusters
as a function of the relative distances between the $\alpha$ clusters.
At each $\alpha$--$\alpha$ distance, the wave function is orthogonalized 
to the ground state. Here the ground state is  represented by the optimal
AQCM basis state ($^4$He--$^4$He distance of $1.4$~fm and $\Lambda = 0.2$) shown by the solid circle. 
Therefore,
the two-by-two matrix is diagonalized at every point on the horizontal axis. 
It is found that the pure cluster state appears around the excitation energy of
$E_x = 15$~MeV with the relative $\alpha$--$\alpha$ distance of $\sim$2.5~fm.
To simplify the discussion, the positions for the Gaussian center parameters for the $\Lambda$ particles
are set to origin only in Figs.~\ref{pure-cluster}~(a) and (b).
\par 
This situation is quite different if the $\alpha$ cluster is assumed to be not broken due to the spin-orbit
interaction in the ground state.  
This is an artificial calculation, but we can clearly see the influence of the cluster-shell competition
in the excited state;
Fig.~\ref{pure-cluster}~(b) shows the result
when the ground state is represented by the Brink model,
which is prepared by changing the $\lambda$ value to zero and the $^4$He--$^4$He distance to 2.2~fm.
The excited $0^+$ state is quite influenced by this change of the ground state.
The energy is pushed up by more than 10~MeV, and the optimal $\alpha$--$\alpha$ distance
is increased to $\sim$3~fm.
This is because if the ground state is a pure three-$\alpha$ cluster state,
the excited states need to be more clusterized to satisfy the orthogonal condition.
On the other hand, if the ground state has different components other than the cluster structure,
it is easier for the pure cluster state to be orthogonal to the ground state.
This effect has been known in $\nuc{C}{12}$ and called the ``shrink effect'' of the second $0^+$ state;
when the $\alpha$ breaking component is mixed in the ground state,
the second $0^+$ state orthogonal to the ground state shrinks.
We found that this shrinking effect is much more enhanced in $^{14}_{\Lambda \Lambda}$C.

\section{Conclusions} 
\label{Concl}
\par
The effect of adding hyperon(s) in nuclear systems is a fundamental problem in nuclear structure physics.
We analyzed this effect in the context of cluster-shell competition and discussed the difference
between Be and C cases.
The antisymmetrized quasi-cluster model (AQCM) is a useful tool to treat the cluster states and shell-model states
on the same footing, and we added $\Lambda$ particle(s) to $^8$Be and $^{12}$C.
\par
The cluster breaking effect is negligibly small in $\nuc{Be}{8}$, where $\alpha$--$\alpha$ cluster structure keeps
enough distance; they stay out of the interaction range of the spin-orbit interaction, 
which breaks the $\alpha$ clusters.
The situation holds even after $\Lambda$ particle(s) is added.
The glue-like effect of $\Lambda$ particles surely shrinks the cluster-cluster distance,
but clusters are not yet broken.
\par
The situation is completely different in the C case since the additional $\alpha$ cluster
shrinks the cluster-cluster distance, and clusters are in the interaction range of the spin-orbit interaction.
The ground state of $\nuc{C}{12}$ contains the component of the $jj$-coupling shell model.
The energy difference between the traditional Brink model and AQCM is about 3.3~MeV in $\nuc{C}{12}$, and this is much enhanced with the increasing number of the $\Lambda$ particles added. 
The energy difference is about 5.2~MeV in $^{14}_{\Lambda \Lambda}$C.
This is because the spin-orbit interaction works in the inner region of the nuclear systems, and the glue-like effect of $\Lambda$ particles shrinks the system and induces more contribution of the spin-orbit interaction.
In $^{14}_{\Lambda\Lambda}$C, the breaking of $\alpha$ clusters in $^{12}$C is much enhanced 
by the addition of the $\Lambda$ particles.
The energy and structure of the excited $0^+$ state with a pure cluster structure are found to be drastically affected
by the transition of the ground state to the $jj$-coupling shell model side.

\begin{acknowledgments}
  This work was supported by JSPS KAKENHI Grant Number 19J20543, 22K03618, and JP18H05407.
  The numerical calculations have been performed using the computer facility of 
  Yukawa Institute for Theoretical Physics,
  Kyoto University (Yukawa-21). 
\end{acknowledgments}
%
\bibliography{hyper.bib}
%
\end{document}